\documentclass[a4paper,twocolumn, superscriptaddress]{revtex4} 
\usepackage{graphicx}

\begin{document}

\title{Vacuum Rabi splitting and strong coupling dynamics for surface plasmon polaritons and Rhodamine 6G molecules}
\author{T.~K.~Hakala}
\affiliation{Nanoscience Center, Department of Physics, P.O. Box 35, FI-40014 University of Jyv\"askyl\"a, Finland}
\author{J.~J.~Toppari}
\affiliation{Nanoscience Center, Department of Physics, P.O. Box 35, FI-40014 University of Jyv\"askyl\"a, Finland}
\author{A. Kuzyk}
\affiliation{Nanoscience Center, Department of Physics, P.O. Box 35, FI-40014 University of Jyv\"askyl\"a, Finland}
\affiliation{Department of Applied Physics, P.O. Box 5100, FI-02015 Helsinki University of Technology, Finland}
\author{M. Pettersson}
\affiliation{Nanoscience Center, Department of Chemistry, P.O. Box 35, FI-40014, University of Jyv\"askyl\"a, Finland}
\author{H. Tikkanen}
\affiliation{Nanoscience Center, Department of Chemistry, P.O. Box 35, FI-40014, University of Jyv\"askyl\"a, Finland}
\author{H. Kunttu}
\affiliation{Nanoscience Center, Department of Chemistry, P.O. Box 35, FI-40014, University of Jyv\"askyl\"a, Finland}
\author{P.~T\"orm\"a}
\affiliation{Department of Applied Physics, P.O. Box 5100, FI-02015 Helsinki University of Technology, Finland}

\begin{abstract}

We report on strong coupling between surface plasmon polaritons (SPP) and Rhodamine 6G
(R6G) molecules, with double vacuum Rabi splitting energies up to 230 and 110 meV. In addition,
we demonstrate the emission of all three energy branches of the strongly coupled SPP-exciton
hybrid system, revealing features of system dynamics that are not visible in conventional reflectometry. 
Finally, in analogy to tunable-Q microcavities, we show that the Rabi splitting can be controlled by adjusting
the interaction time between waveguided SPPs and R6G deposited on top of the waveguide. The
interaction time can be controlled with sub-fs precision by adjusting the length of the R6G area
with standard lithography methods.

\end{abstract}
\maketitle
In the strong coupling regime, the Fermi Golden Rule fails and coherent dynamics dominates. For light, strong coupling is important, e.g., in lasing and in coherent energy transfer between excitations, relevant for light-harvesting systems. Vacuum Rabi splitting (VRS) has been observed in microcavities for inorganic \cite{Weisbuch,Khitrova} and organic \cite{Lidzey, Holmes} semiconductors. Surface plasmon polaritons (SPP) provide an alternative route to enhance the coupling and extend to nanoscale. Rabi splitting between SPP and J-aggregates, having particularly narrow absorption linewidth, was demonstrated in \cite{Bellessa, Dintinger, Pockrand}. Here we show a double VRS between SPP and R6G molecules with splitting energies of 230 meV and 110 meV. 
Further, we observe the emission of all three energy branches of the strongly coupled system, which is promising concerning light generation and can be used to probe the system dynamics, such as energy transfer between the three exciton-SPP branches. Besides using Kretschmann geometry \cite{Kretschmann, Zayats}, we study dynamics of strong coupling also with waveguided SPPs propagating through a molecular area of controllable length, in analogy with a tunable-Q cavity. 
Additionally, the observation of Rabi split, despite the broad absorption linewidth of R6G, suggests usability of a wide variety of organic molecules for achieving strong coupling. 



To study the strong coupling between SPPs and R6G, sandwich structured samples were fabricated, having a glass substrate, a 45 nm silver film above it, and on the top a resist layer (50 nm, Microchem SU-8 2025) with four different concentrations of R6G. To determine the absorbance of the molecular films, reference samples with no silver layer were fabricated. For the SPP waveguide studies, another set of samples were fabricated (for details, see \cite{Kuzyk, Hakala}), allowing spectral measurements in specific locations on the sample and the control of the interaction time between R6G and SPP. 

\begin{figure}[htb]
\includegraphics[width=70truemm]{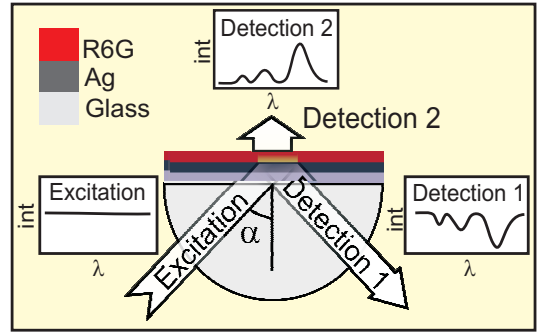}
\caption{The samples were studied by using two different, complementary detection methods. Detection 1 (DM1) gives the energies that are coupled into the system as a function of in-plane wave vector k. The coupled modes are shown as dips in the reflected light. Detection 2 (DM2) gives the energies of the modes that are coupled out from the system by scattering processes. }
\label{fig:setup}
\end{figure}

Reflectometry measurements in the Kretschmann configuration were performed [see Fig.~\ref{fig:setup}, detection 1 (DM1)], yielding the dispersion relation of the energy of incoupled modes as a function of the in-plane wave vector k. Another method was used for studying the dynamics and mode emission after the incoupling [see Fig.~\ref{fig:setup}, detection 2, (DM2)]. In this case we employ the scattering of the strongly coupled hybrid modes into photons from scattering centers always present in practical systems (silver impurities or small corrugations).  The reflected (DM1) or luminescent (DM2) light was collected by focusing the light into an optical fiber connected to a spectrometer. 
For the excitation, a collimated, p-polarized white light source was used. The incident angle was controlled by a rotatable prism. 

\begin{figure}[htb]
\includegraphics[width=75truemm]{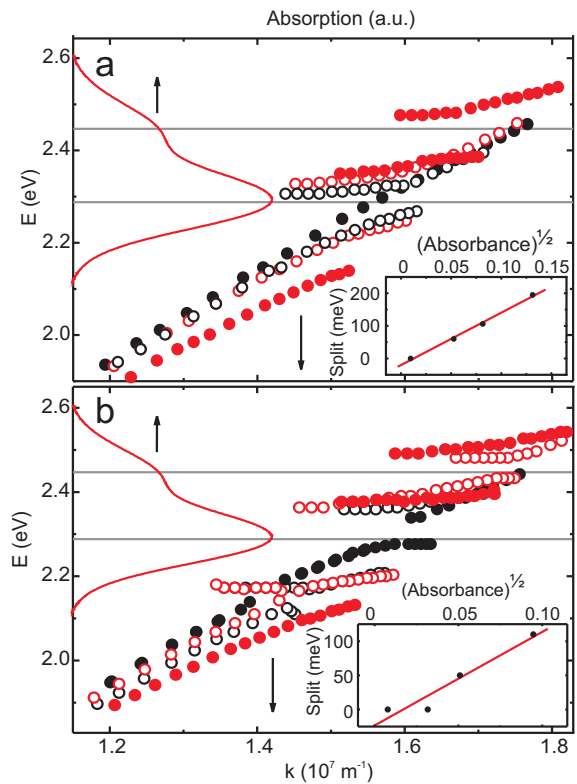}
\caption{The dispersions of four different samples having R6G concentrations of 4 mM (solid black circles), 25 mM (empty black circles), 50 mM (empty red circles) and 200 mM (solid red circles) measured by DM1 (a) and DM2 (b). The solid red curves are the measured absorbance of the 200 mM SU-8/R6G film, and solid grey lines are the absorption maximum and absorption shoulder energies. The insets show the low (a) and high (b) energy splits as a function of (absorbance)$^{1/2}$ with linear fits.}
\label{fig:dispersions}
\end{figure}

The dispersions measured with DM1 are shown in Fig.~\ref{fig:dispersions}a. For the 4 mM concentration, the dispersion has only one mode which is characteristic to a SPP on a silver/SU-8/air -structure, i.e., no molecular contribution. When increasing the concentration to 25 mM, an anticrossing behaviour emerges at the energy corresponding to a measured absorption maximum of R6G in SU-8 (2.29 eV) as a sign of the strong coupling regime. By increasing the concentration to 50 mM, this Rabi splitting widens. When the R6G concentration is further increased to 200 mM, the existing split again widens and a second split appears at around 2.45 eV, which corresponds to a measured absorption shoulder of the R6G film. The inset of Fig.~\ref{fig:dispersions}a shows the expected linear relationship between the low energy splitting and the square root of absorbance \cite{Skolnick}. 

The Rabi-splitting energies for the 200 mM sample are 200 meV (main absorption) and 100 meV (absorption shoulder). The middle branch thus presents SPP-induced exciton hybridization between the two excitons. Previous observations of exciton hybridization with organic dye molecules have involved cavity photons \cite{Lidzey2, Holmes2}, whereas here the SPP creates the hybridization. 

Previously, the strong coupling regime between SPPs and photoactive organic molecules has been observed only with J-aggregates, with Rabi-splitting energies up to 250 meV \cite{Bellessa, Dintinger, Pockrand, Sugawara, Wurtz, Fofang}. For these aggregates, the Rabi splitting at room temperature has been associated with the high oscillator strength together with a small absorption linewidth \cite{Hobson}. It should be stressed that we observe Rabi splitting although, for each transition (the main and the shoulder), the minimum observed Rabi splitting is below the original width of that transition. This suggests that the SPP-molecule interaction causes line narrowing as a precursor to the strong coupling \cite{Sugawara, Ritchie, Noginov}.

To test the effect of the photon number, we varied the intensity over two orders of magnitude, i.e., between 0.17-13 W/m$^2$ over the visible range, with no effect on the splitting, consistent with VRS. Note that this is the so-called many-atom/quantum well VRS: many molecules are interacting coherently with the same SPP-photon.

Dynamics, that is, energy transfer, propagation and scattering, as well as dissipation mechanisms, of the hybrid modes are essential, yet scarcely studied phenomena in these systems. To study dynamics, we employ DM2 shown in Fig.~\ref{fig:setup}. The spectrum measured with DM2 has negligible contribution from the reflected or transmitted excitation light: it consists of scattered hybrid SPP-exciton modes and uncoupled, spontaneous emission of R6G. Since the scattering event is temporally separated from the incoupling event, we gain insight into the dynamics of the system by comparing the energies of these two events. The excitation being the same both in DM1 and DM2 implies that all differences between the results must come from the dynamics after incoupling, not from the incoupling mechanism. Figure \ref{fig:dispersions}b shows the dispersions for the same samples, measured by DM2. We emphasize that we see emission from all three energy branches, in contrary to previous reports \cite{Bellessa, Symonds, Symonds2}, where the emission from the upper branch was always missing. In contrast to ours, the excitation in these experiments was done from the side of the molecules.

\begin{figure}[htb]
\includegraphics[width=50truemm]{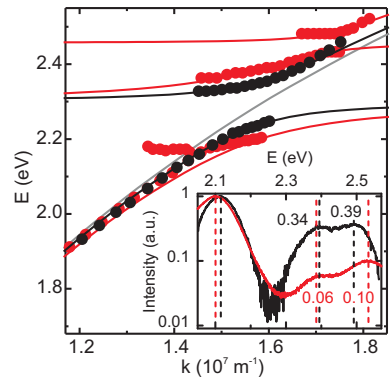}
\caption{The measured dispersions for 50 mM sample for DM1 (black) and DM2 (red) and their theoretical fits. The solid grey curve is the theoretical dispersion of the uncoupled SPP on Ag/SU-8/air structure. The inset shows the normalized spectra for 200 mM sample with DM1 (black) (background subtracted), and DM2 (red), both with excitation light angle $71.5^{\circ}$. The dashed lines show the spectral peak positions for the measurement methods DM1 (black) and DM2 (red). The numbers indicate the intensity of the peaks}
\label{fig:energytransfer}
\end{figure}

One qualitative difference between the DM1 and DM2 dispersions is that DM2 shows an additional emission branch nearly independent of the in-plane k vector at around 2.17 eV, for the 25 and 50 mM samples. According to the measurements of a reference sample having R6G film but no silver, this branch could be identified as the spontaneous, non-coupled, emission maximum of R6G, reported also for J-aggregates \cite{Bellessa, Hobson}. The more significant difference in the dispersions is that the DM2 shows an increased energy splitting as compared to DM1, e.g., 230 meV and 110 meV for the 200 mM sample. Particularly distinctive is the difference for the 4 mM and 50 mM samples, in which the low or high energy splitting, respectively, is apparent only for DM2 (see Fig.~\ref{fig:dispersions}a-b and for clearer comparison Fig.~\ref{fig:energytransfer} with theoretical fits included). 

Although the maximum VRS is proportional to $A \sqrt{N/V}$, where $A$ contains the transition dipole moment, $N$ is the number of oscillators and $V$ the mode volume, dynamics such as decay and decoherence may decrease the Rabi splitting \cite{Agranovich}. For instance, in case of cavities, the VRS is $2 \sqrt{g^2-(\gamma_C -\gamma_X)^2/16}$, where $g$ is a coupling constant and $\gamma_C$ ($\gamma_X$) is the cavity (exciton) decay rate \cite{Reithmaier, Yoshie}. The widths of the modes are given by $(\gamma_C + \gamma_X)/2$. A simple interpretation of the observed larger splitting in DM2 can be suggested using this analogy. In our case, the incoupling is via near-field component of the incident photon: the timescale of the event equals the time the photon spends in the immediate vicinity of the surface. After this time, there is no possibility for the strongly coupled system to affect the signal measured by DM1. In this way, we enforce in DM1 a fast decay, i.e., a small effective interaction time. In contrast, for DM2, the interaction time of SPPs with the molecular film is only limited by the decay or scattering of SPPs. The longer path length in case of DM2 is like the larger number of round trips a photon makes in a higher Q cavity. 

\begin{figure}[htb]
\includegraphics[width=0.45\textwidth]{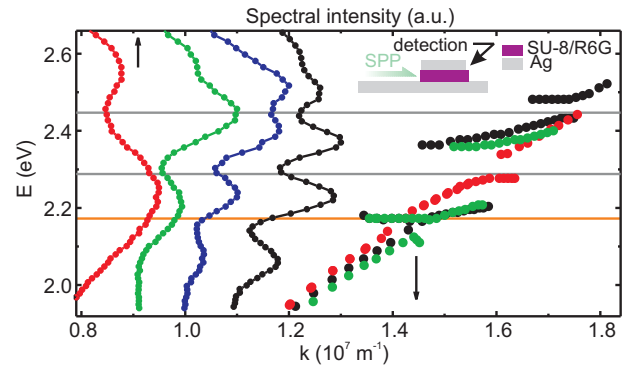}
\caption{The emission spectra measured from the waveguide samples having different lengths of R6G deposited on top of a waveguide (see the inset schematic), along with the dispersion curves of the thin film samples measured with DM2. The red, green and blue spectra correspond to samples having 1 $\mu$m, 2 $\mu$m and 5 $\mu$m lengths of R6G, respectively. The black spectrum is measured from a waveguide sample having a 5 $\mu$m R6G area as well, but with a layer of silver deposited on top this area (as in the schematic). In the dispersions red, green and black correspond to 4, 25 and 50 mM samples, respectively. The orange vertical line is the measured R6G emission and the grey lines absorption maximum and absorption shoulder energies.}
\label{fig:waveguide}
\end{figure}

The above reasoning was tested by fabricating plasmonic silver waveguides where SPPs are launched at a certain location and, after propagating micrometers, go through an interaction area of R6G molecules in an SU-8 matrix on top of the waveguide. The emitted spectrum at the end of the interaction area was recorded. The launched SPPs have a broad spectrum peaked at 2.55 eV, overlapping with the studied R6G transitions. For the fabrication and measurements of such structures see our previous publications \cite{Kuzyk, Hakala}. By limiting the length of the interaction area, we wanted to enforce a faster effective decay than given by the natural coherence times/lengths of the system. One may assume the SPP coherence length to be similar to the propagation length of SPP \cite{Ditlbacher} (5 $\mu$m in our case \cite{Kuzyk}): we studied samples with the interaction area lengths of 1, 2 and 5 $\mu$m. As seen in Fig.~\ref{fig:waveguide}, for the shortest area, only a broad peak around the R6G absorption and emission is visible (2.25 eV) \cite{shift}.  The other peak (2.55 eV) is just due to the incoming non-coupled SPPs. For the 2 $\mu$m sample, a clear splitting is seen around the main absorption, and the peaks are narrowed from the 1 $\mu$m case. For 5 $\mu$m, also the second split at the absorption shoulder appears. Moreover, there is an additional feature possibly related to the R6G emission around 2.17 eV, indicating that also this transition is approaching the strong coupling regime. Apart by the interaction area length, we increased the SPP-matter interaction by fabricating a layer of silver on top of the interaction area, thereby decreasing the mode volume and blocking the decay into radiative modes. With such a sample, the strong coupling features become very prominent.  

Finally, we consider the possible energy transfer between the hybrid modes \cite{Agranovich2, Chovan}. The inset of Figure \ref{fig:energytransfer} shows the normalized spectra of the 200 mM sample measured with both DM1 and DM2, with $71.5^{\circ}$ incident angle. The relative intensity of the two high energy modes is lower for DM2. The highest energy mode has 1/4 and the middle one 1/6 intensity in DM 2 as compared to DM1. The differences originate from the time evolution of the system: propagation speed, dissipation, and scattering properties of the modes, and/or energy transfer between the modes. Dissipation and scattering are nearly wavelength independent for the values considered \cite{Hakala}, which indicates that energy transfer plays a considerable role. To confirm the existence of the energy transfer is, however, beyond the scope of this letter. 

It should be noted that the high molecular concentration of our samples together with the tendency of R6G to form aggregates, particularly dimers \cite{Levshin, Bojarski}, implies that the possible dimer contribution to the energy splits cannot be neglected. The R6G dimer absorption in solid matrices forms two bands, H and J: the absorbance maximum of the H band overlaps with the monomer shoulder, whereas the absorbance of the J band is strongly redshifted with respect to monomer maximum \cite{Bojarski}. Furthermore, the dimer has virtually zero absorbance at the energy corresponding to monomer absorption maximum. Thus, the aggregate percentage in our experiments can be deduced from the relative increase of the absorption shoulder with respect to the maximum. For this, the absorbances of the reference samples with the same parameters as the samples under study, but with no silver layer, were  measured with a spectrometer and compared with the literature values of monomer R6G \cite{Bojarski}. 
From this data we estimated the fraction of absorbance originating from aggregates, for each sample, to be 0\%, 0\%, 15\% and 65\% for the 4 mM, 25 mM, 50 mM and 200 mM samples, respectively. 

Since the low energy split in our samples is located at the monomer absorption maximum rather than the dimer J band maximum, we conclude that this split originates from the R6G monomers. 
The high energy split is visible with relative small aggregate concentrations (15\%), therefore we expect that both the monomers and dimers contribute to that splitting. In addition, since the dimer J band overlaps with the monomer emission, we cannot totally exclude the contribution of this to the features around 
2.17 eV in the DM2 and waveguide experiments, see Figs. \ref{fig:dispersions}-\ref{fig:waveguide}. However, due to J band being weaker than the H band and the small amount (15\%) of aggregates implies the emission around 2.17 eV being mostly due to monomer emission. The observed emission around 2.17 eV is likely to have contribution from uncoupled emission, however, as seen from Figs \ref{fig:dispersions}b and \ref{fig:energytransfer}, it is not totally independent of the in-plane k vector but rather has a small tilt following the SPP dispersion curve. This could be an indication of the emergence of strong coupling regime also for this transition.

In summary, we have demonstrated a double VRS for SPP and R6G molecules, despite the broad absorption of R6G. We therefore expect that the strong coupling regime between SPPs and photoactive molecules is readily accessible for a wide variety of other molecules as well. For the first time, we demonstrate that each of the three energy branches of the strongly coupled SPP-molecule system can be converted into photons. 
Especially, {\it dynamics} was studied by comparison of the energies and intensities of the in- and outcoupled modes in reflectometry, and by a novel approach using waveguide experiments. In the latter case, the effective Rabi splitting was easily controlled by standard lithographical methods which is promising considering applications. The double split, as well as the potential of creating spatially separated interaction areas on the waveguides, open up interesting possibilities for studies of multimode hybridization and energy transfer.  

{\it Acknowledgements} This work was supported by the Academy of Finland (Project Nos. 117937, 118160, 115020, 213362) and conducted as part of a EURYI scheme award. See www.esf.org/euryi. The authors thank Pasi Myllyperki\"o and Klas Lindfors for fruitful discussions. A.K. thanks the National Graduate School in Nanoscience.


\begin{thebibliography}{99}
\bibitem{shift} The spectra in the Fig.~\ref{fig:waveguide} are shifted approximately 9 nm (40 meV) to compensate for 
instrument errors in the confocal microscopy. Even without this correction the spectra show high qualitative agreement with the measured dispersion curves.
\end{thebibliography}
\end{document}